\documentclass[showpacs,prb,manuscript,aps]{revtex4}
\usepackage{graphicx}
\usepackage{amsmath}
\begin{document}
\title{Phonon spectra and superconductivity of BiS$_2$-based
compounds LaO$_{1-x}$F$_{x}$BiS$_2$}
 
\author{B. Li}
\author{Z. W. Xing}
\email{zwxing@nju.edu.cn}
\affiliation{National Laboratory of Solid
State Microstructures and Department of Materials Science and
Engineering, Nanjing University, Nanjing 210093, China}
\author{G. Q. Huang}
\affiliation{Departement of Physics, Nanjing Normal University, Nanjing
210046, China}

\begin{abstract}{
Using the density-functional perturbation theory with structural
optimization, we investigate the electronic structure, phonon
spectra, and superconductivity of BiS$_2$-based layered compounds
LaO$_{1-x}$F$_{x}$BiS$_2$. For LaO$_{0.5}$F$_{0.5}$BiS$_2$, the
calculated electron-phonon coupling constant is equal to
$\lambda=0.8$,  and obtained $T_c \simeq$ 9.1 K is very close to its
experimental value, indicating that it is a conventional
electron-phonon superconductor.}
\end{abstract}

\maketitle .

The recent discovery of superconductivity in the BiS$_2$-based
compounds has attracted much attention due to their layered crystal
structure and physical properties similar to cuprate and Fe-based
superconductors. The Bi$_4$O$_4$S$_3$ superconductor was found to
exhibit metallic transport behavior and to show a zero-resistivity
state below 4.5 K.~\cite{1} Soon after, another type of
BiS$_2$-based systems, ReO$_{1-x}$F$_{x}$BiS$_2$ (Re = La, Ce, Pr,
and Nd), were reported to be superconducting with transition
temperatures $T_c$ equal to 10.6, 3.0, 5.5, and 5.6 K,
respectively.~\cite{2, 3, 4, 5} For Bi$_4$O$_4$S$_3$
(ReO$_{1-x}$F$_{x}$BiS$_2$), superconductivity can be obtained by
electrons doping into the insulating parent compound
Bi$_6$O$_8$S$_5$ (ReOBiS$_2$). Experimental and theoretical studies
indicated that these materials exhibit multiband behavior with
dominant electron carriers originating mainly from the Bi $6p$
orbitals.~\cite{1, 6, 7, 8, 9, 10, 11, 12}
\par

In this Letter we report the calculated results for the electronic
structure of LaO$_{1-x}$F$_{x}$BiS$_2$ ($x =$ 0, 0.5, and 1), phonon
dispersions and electron-phonon coupling of superconducting
LaO$_{0.5}$F$_{0.5}$BiS$_2$. From the first-principles calculation,
it is shown that the parent compound LaOBiS$_2$ is a band insulator
with an energy gap of $\sim$ 0.8 $eV$, similar to the ground state
of cuprate and pnictide superconductors. Upon F doping, the Fermi
level moves up and the system appears a metallic nature. We obtain
electron-phonon coupling constant $\lambda=0.8$ and superconducting
transition temperature $T_c \simeq$ 9.1 K in
LaO$_{0.5}$F$_{0.5}$BiS$_2$, indicating that this compound can be
explained as a conventional phonon-mediated superconductor, quite
different from the Fe-based materials~\cite{13, 14} but somewhat
similar to the Ni-based compounds.~\cite{15, 16}

\par
We used the full-potential linearized augmented plane wave method
implemented in the WIEN2K package~\cite{17} for the lattice
parameter optimization and electronic structure calculations. The
generalized gradient approximation (GGA) presented by Wu and
Cohen~\cite{18} was used for the exchange-correlation energy
calculations. This GGA is a nonempirical approximation that gives
significant improvements of calculations for lattice constants and
crystal structures. The phonon dispersion and the electron-phonon
coupling are calculated using density functional perturbation theory
(DFPT). The phonon dispersions were obtained in linear response via
the QUANTUM-ESPRESSO program~\cite{19} and ultrasoft
pseudopotentials. The GGA of Perdew-Burke-Ernzerhof~\cite{20} was
adopted for the exchange-correlation potentials with a cutoff of 60
Ry for the wave functions and 600 Ry for the charge density.

\par
LaO$_{1-x}$F$_{x}$BiS$_2$ with $x$=0 or 1 is crystallized in the
ZrCuSiAs type tetragonal structure (P4/nmm) with La, Bi, S1 in the
Bi planes, S2 in between Bi-S1 and La-O(F) planes, and O(F) being at
the positions $2c$(0.5, 0, $z_{La}$), $2c$(0.5, 0, $z_{Bi}$),
$2c$(0.5, 0, $z_{S1}$), $2c$(0.5, 0, $z_{S2}$), and $2a$(0, 0, 0),
respectively. The F substitution at the O sites in
LaO$_{0.5}$F$_{0.5}$BiS$_2$ reduces the symmetry to a space group of
P-4m2. Starting from the lattice parameters  ($a =$ 4.0527 $\AA$ and
$c =$ 13.3237 $\AA$) reported by Mizuguchi $et$ $al.$~\cite{2}, we
have performed the full structural optimization including the
lattice parameters and atomic positions, the obtained results being
summarized in Table \ref{tab.1}. It is found that, with doping F from $x=0$ to
$x=1$, the $a$ value increases gradually but the $c$ value
decreases, accompanied with a decrease of primitive cell volume $V$.
By making a comparison in the atomic bond length between
LaO$_{0.5}$F$_{0.5}$BiS$_2$ and LaOBiS$_2$, one finds that the Bi-S2
bond expands due to F doping while $d_{Bi-S1}$ for the Bi-S1 bond
remains almost unchanged.

\begin{table}
\caption{Various optimized structural parameters for
LaO$_{1-x}$F$_{x}$BiS$_2$ with $x$=0, 0.5, and 1.}\label{tab.1}
\begin{center}
\begin{tabular}{c c c c }
\hline
    & LaOBiS$_2$ & LaO$_{0.5}$F$_{0.5}$BiS$_2$ & LaFBiS$_2$ \\
  \hline
  $a$ (\AA)       & 4.0677   &4.1091  &4.1524 \\
  $c$ (\AA)       & 14.3085  &13.4196 &12.9554\\
  $V$ (\AA$^3$)   & 236.75   &226.59  &223.38 \\
  $z_{La}$        & 0.0890   &0.1073  &0.1276 \\
  $z_{Bi}$        & 0.6309   &0.6145  &0.6090\\
  $z_{S1}$        & 0.3945   &0.3844  &0.3691\\
  $z_{S2}$        & 0.8073   &0.8128  &0.8163\\
  $d_{Bi-S1}$(\AA)& 2.8993   &2.9056  &2.9499\\
  $d_{Bi-S2}$(\AA)& 2.5231   &2.6621  &2.6862\\
  \hline
\end{tabular}
\end{center}
\end{table}

\par
\begin{figure*}
\begin{center}
\includegraphics[width=8cm]{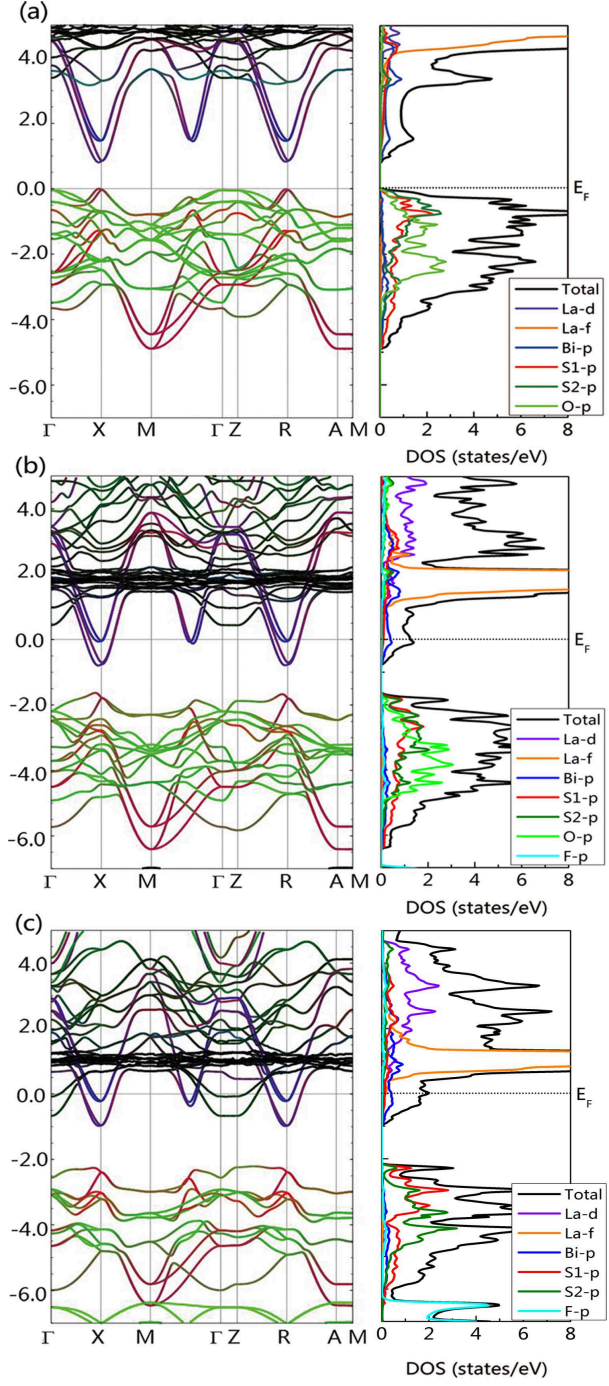}
\caption{Calculated band structure and density of states for
LaO$_{1-x}$F$_{x}$BiS$_2$
  with $x$=0 (a), 0.5 (b), and  1 (c) with orbital character
  indicated by different colors: Bi-$p$ (blue), S1-$p$ (red),
  and O-$p$ and S2-$p$ (green).} \label{fig.1}
\end{center}
\end{figure*}

The calculated band structures and electronic density of states
(DOS) are shown in Fig. \ref{fig.1}. The contribution of orbital states to
the band structure are characterized by different colors: blue
(Bi-$p$), red (S1-$p$), and green (O-$p$ and S2-$p$). For
LaOBiS$_2$, the Fermi level is located at the upper edge of valence
bands with an energy gap of $\sim$ 0.8 eV, indicating an insulating
behavior. The valance bands spread from around $-6$ eV to 0  and
consist of the $p$ states of the O and S atoms.  While the La-$d$
and La-$f$ states lie above 4 eV, far away from the Fermi level, the
Bi-$p$ and S1-$p$ states dominate the conduction bands. For doped
LaO$_{0.5}$F$_{0.5}$BiS$_2$, the Fermi level enters the conduction
bands coming from the Bi-$p$ and S1-$p$ states, so that the Bi-S1
layers dominantly contribute to the electronic conduction. As shown
in Fig. \ref{fig.1}b, the present Fermi level crosses four bands of electron
pockets, and so LaO$_{0.5}$F$_{0.5}$BiS$_2$ has electron carriers,
which is well consistent with the Hall resistance
measurements.~\cite{10} The DOS at the Fermi level is equal to
$N(E_{F})$ = 1.22 eV$^{-1}$ on a per formula unit both spins basis,
and the corresponding bare susceptibility and specific heat
coefficient are given by  $\chi_0 = 4.0 \times 10^{-5}$ emu/mol and
$\gamma_0$ = 3.0 mJ/mol K$^2$, respectively. For LaFBiS$_2$, the
Fermi level further enters the conduction bands and the energy gap
increases to $\sim$ 1.1 eV. Meanwhile, the DOS at the Fermi level
increases to $N(E_{F})$ = 1.84 eV$^{-1}$ per formula unit both
spins, yielding $\chi_0 = 6.0 \times 10^{-5}$ emu/mol and $\gamma_0$
= 4.5 mJ/mol K$^2$.

\begin{table}
\caption{Phonon mode frequencies (cm$^{-1}$) at $\Gamma$ and $M$
points in LaO$_{0.5}$F$_{0.5}$BiS$_2$.
   I: infrared active, R: Raman active.}\label{tab.2}
\begin{center}
\begin{tabular}{c c c c c c c c c c c c c c c c c}
\hline
&&&$\Gamma$(0, 0, 0)&&&\\
\hline
  $B_2(I+R)$& 0& 65.4& 145.4&249.4& 270.1&412.2\\
  $E(I+R)$       & 0& 39.3& 44.0& 78.4& 122.8\\
            & 153.6& 171.3&184.1& 243.8&345.1\\
  $A_1(R)$     & 74.8& 138.3&168.4 & 305.9\\
  \hline
  &&&$M$(0.5, 0.5, 0)&&&\\
\hline
  $A_1$     & 61.3& 81.3&155.8&195.8\\
  $A_2$     & 55.5&101.7&123.5&189.3&275.8\\
  $B_1$     & 60.8&81.6&169.5& 199.6\\
  $B_2$     & 56.3&94.3&103.3&191.0&299.6\\
  $E$       & 71.0&74.9&148.4&220.5&269.8&403.4\\
\hline
\end{tabular}
\end{center}
\end{table}

\par
\begin{figure*}
\begin{center}
\includegraphics[width=10cm]{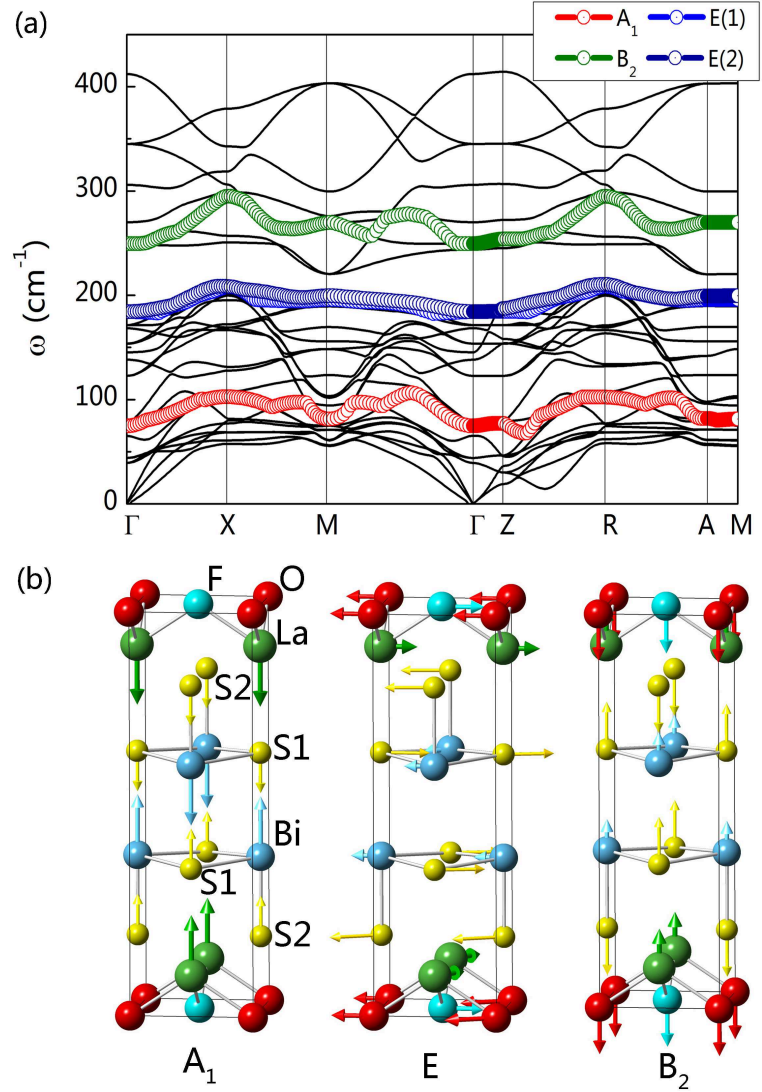}
\caption{(a) Calculated phonon dispersion curves of
  LaO$_{0.5}$F$_{0.5}$BiS$_2$. (b)
  Atom vibration pattern of the $A_1$, $E$ and $B_2$ modes at $\Gamma$ point.} \label{fig.2}
\end{center}
\end{figure*}
\begin{figure*}
\begin{center}
\includegraphics[width=10cm]{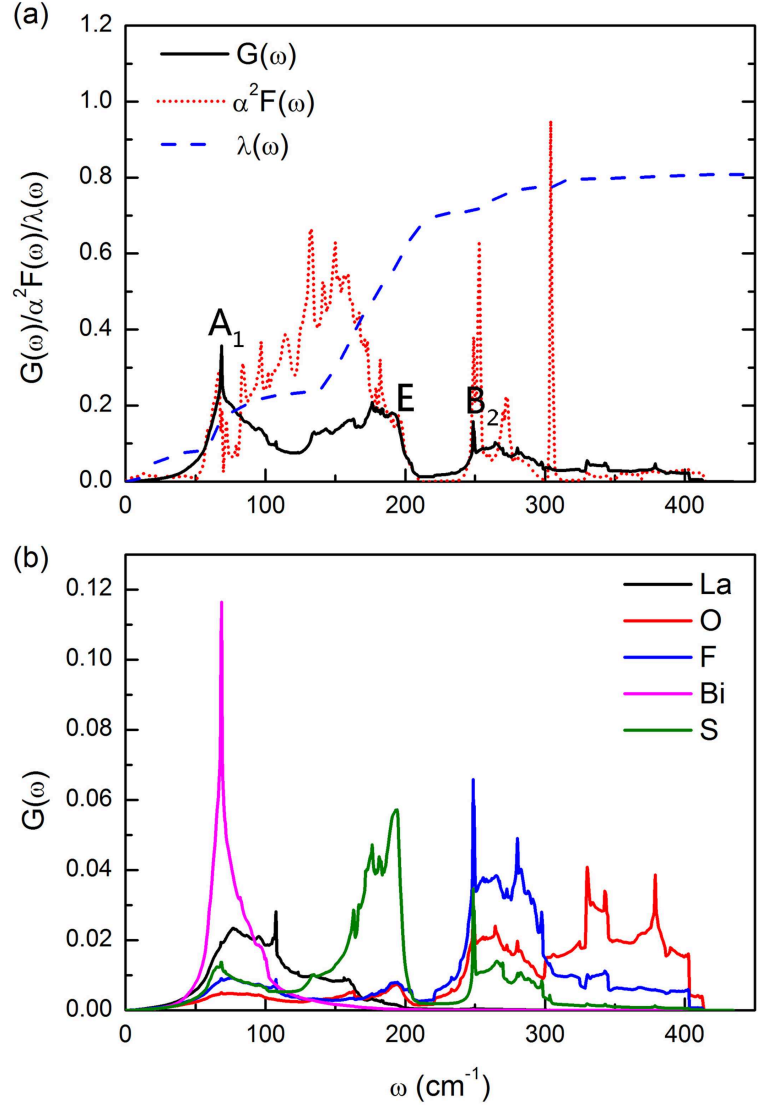}
 \caption{ (a) Phonon density of states G($\omega$),
  electron-phonon spectral function $\alpha^2F(\omega)$ and electron-phonon coupling
  $\lambda(\omega)$ for LaO$_{0.5}$F$_{0.5}$BiS$_2$. (b) Projected phonon density of
  states for each atom.} \label{fig.3}
\end{center}
\end{figure*}

The calculated phonon dispersions of LaO$_{0.5}$F$_{0.5}$BiS$_2$ are
plotted in Fig. \ref{fig.2}a, in which there are 30 phonon bands
extending up to $\sim$ 450 cm$^{-1}$ and the point group at the
$\Gamma$ and $M$ points is $D_{2d}$. The $\Gamma$ and $M$-point
modes can be decomposed as $\Gamma=6B_2\bigoplus10E\bigoplus4A_1$
and $M=4B_1\bigoplus4A_1\bigoplus5B_2\bigoplus5A_2\bigoplus6E$. Note
that the  $E$ mode is degenerate, and its frequencies are listed in
Table \ref{tab.2}. The Raman modes at the zone center can be
measured directly by experiments in near future. The corresponding
phonon density of states (PDOS), Eliashberg spectral function
$\alpha^2F(\omega)$, electron-phonon coupling $\lambda(\omega)$ and
projected PDOS are shown in Fig. \ref{fig.3}. There are three
distinct peaks centered at $\sim$ 70 cm$^{-1}$ ($A_1$ mode at
$\Gamma$ point), $\sim$ 190 cm$^{-1}$ ($E$ mode at $\Gamma$ point),
and $\sim$ 250 cm$^{-1}$ ($B_2$ mode at $\Gamma$ point),
respectively, corresponding to three thick lines from lower to upper
in Fig. \ref{fig.2}a. As shown in Fig. \ref{fig.2}b, the $A_1$ mode
corresponds to the vertical vibrations with the upper and lower
BiS$_2$-layers (La-layers) beating against each other. The $E$ mode
corresponds to the in-plane stretching vibrations with motion along
the $x(y)$ axis.  In the $B_2$ mode, the La, Bi, and S1 atoms
vibrate along the same $z$ direction, opposite to the vibration
direction of the S2, O, and F atoms. From Fig. \ref{fig.3}, one can
see that the bands around 70 cm$^{-1}$ comes mainly from the PDOS
contributions of the Bi and La motion, the peak at 190 cm$^{-1}$
mainly corresponds to the S character, and that at 250 cm$^{-1}$
corresponds to F, O and S character. The calculated Eliashberg
spectral function has peaks in consistent with the PDOS at the
low-frequency regime, and is evidently enhanced in the intermediate
frequency regime. In the DFPT calculations, the Eliashberg spectral
function depends directly on the electron-phonon matrix element:
\begin{equation}
\begin{split}
\alpha^{2}F(\omega)=\frac{1}{N(E_F)N_{k}}\sum_{kq\nu}
    \mid g^{\nu}_{n\mathbf{k},m(\mathbf{k}+\mathbf{q})}\mid ^{2}\\
    \times\delta(\varepsilon_{n\mathbf{k}})
    \delta(\varepsilon_{m(\mathbf{k}+\mathbf{q})})\delta(\omega-\omega_{q\nu}).
\end{split}
\end{equation}
Here, $N_{k}$ is the number of k points used in the summation,
$N(E_F)$ is the density of states at the Fermi level, and
$\omega_{g\nu}$ are the phonon frequencies. The electron-phonon
matrix element $\mid
g^{\nu}_{n\mathbf{k},m(\mathbf{k}+\mathbf{q})}\mid ^{2}$ is defined
by the variation in the self-consistent crystal potential. It
follows that the  electron-phonon interaction of Bi-S1 bond plays an
important role in the enhancement of $\alpha^{2}F(\omega)$. From the
$\lambda(\omega)$ curves in Fig. \ref{fig.3}a, where
$\lambda(\omega)=2\int^{\omega}_{0}[\alpha^2F(\Omega)/\Omega]d\Omega$,
one finds that $\lambda(220)\simeq 0.7$ has been close to $\lambda
(\infty)\simeq 0.8$, indicating that the phonon modes in the low and
intermediate frequency regimes have the main contribution to the
electron-phonon coupling. From the $\alpha^2F(\omega)$ spectra in
Fig. \ref{fig.3}, we get the electron-phonon coupling  $\lambda$ =
0.8 and the logarithmically averaged frequency $\omega_{ln}$ =192 K.
Using further the Allen-Dynes formula with the Coulomb parameter
$\mu^*$= 0.1, we finally obtain  $T_c$ = 9.1 K, which is found  very
close to the experimental value $T_c$ = 10.6 K.~\cite{2} It then
follows that the doped LaO$_{0.5}$F$_{0.5}$BiS$_2$ is very likely a
conventional electron-phonon superconductor.

\par
In summary, the present first-principles calculations indicate that
LaO$_{1-x}$F$_{x}$BiS$_2$ changes from a band insulator at $x=0$ to
a band metal upon doping. Similar to LaFeAsO$_{1-x}$F$_{x}$,
electrons in LaO$_{1-x}$F$_{x}$BiS$_2$ are doped from the LaO(F)
redox layer to the superconducting BiS$_2$ layers. The $p$-orbital
states of Bi-S1 layers play an important role in the electron
transformation due to their domination of the DOS at the Fermi
level. From the phonon calculations for LaO$_{0.5}$F$_{0.5}$BiS$_2$,
we obtain a strong electron-phonon coupling of $\lambda$ = 0.8 and a
superconducting transition temperature of $T_c$ = 9.1 K, the latter
being very close to its experimental value. It then follows that
LaO$_{0.5}$F$_{0.5}$BiS$_2$ is a conventional electron-phonon
superconductor.
\par
For LaO$_{1-x}$F$_{x}$BiS$_2$, we have shown that the electronic
states near the Fermi level are dominantly from the hybridized $p$
orbital states of the Bi and S1 atoms, and the electron-phonon
coupling comes mainly from the contribution of the Bi and S phonons.
These facts indicate that the BiS$_2$ layers play an important role
in the transport and superconducting properties, similar to the
CuO$_2$ layers of the cuprates or Fe-As layers of the Fe-based
superconductors.

\section{Acknowledgments}
This work is supported by the State Key Program for Basic Researches
of China (2010CB923404), the National Natural Science Foundation of
China (11074109 and 11174125), the National "Climbing" Program of China (91021003),
and the Natural Science Foundation of Jiangsu Province (BK2010012).\\

\newpage

\vskip 0.5in
\begin{center}{\bf Figure Caption}
\end{center}\ \par

Figure 1. (Color online) Calculated band structure and density of
states of LaO$_{1-x}$F$_{x}$BiS$_2$  for $x$=0 (a), 0.5 (b), and 1
(c), with orbital character indicated by different colors: Bi-$p$
(blue), S1-$p$ (red), and O-$p$ and S2-$p$ (green).\\

Figure 2. (Color online) (a)Calculated phonon dispersion curves of
LaO$_{0.5}$F$_{0.5}$BiS$_2$. (b)Atom vibration pattern of the $A_1$, $E$ and $B_2$ modes
at $\Gamma$ point. \\

Figure 3. (Color online) (a) Phonon density of states G($\omega$),
  electron-phonon spectral function $\alpha^2F(\omega)$ and electron-phonon coupling
  $\lambda(\omega)$ for LaO$_{0.5}$F$_{0.5}$BiS$_2$. (b) Projected phonon density of
  states for each atom.\\

\newpage

\end{document}